\documentclass[12pt,a4paper]{article}
\usepackage{amsmath,amssymb}
\usepackage{graphicx}
\usepackage{cite}

\newsavebox{\boxa}
\sbox{\boxa}{\includegraphics[height=2.5cm]{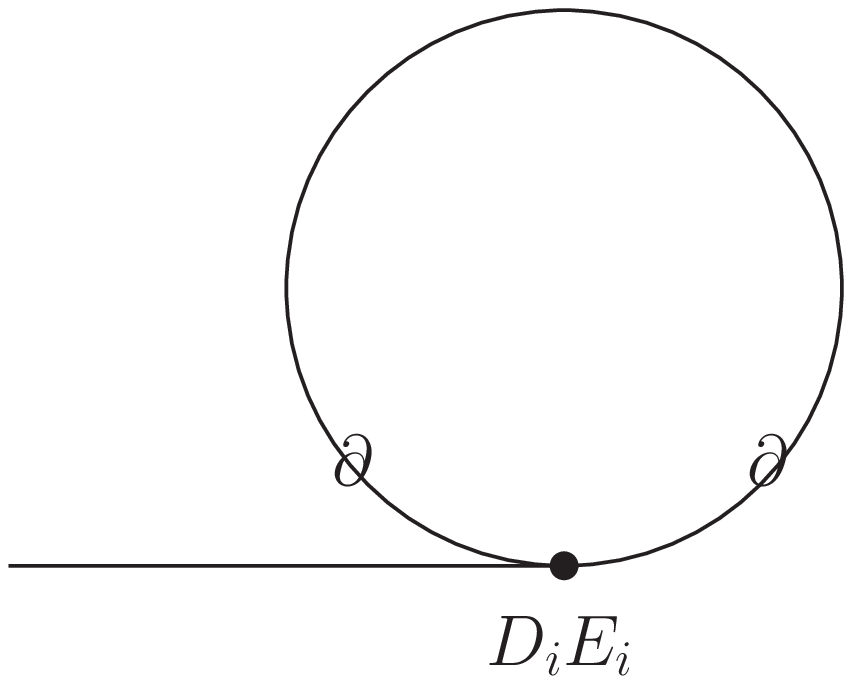}} 
\newlength{\boxal}
\newsavebox{\boxb}
\sbox{\boxb}{\includegraphics[height=2.5cm]{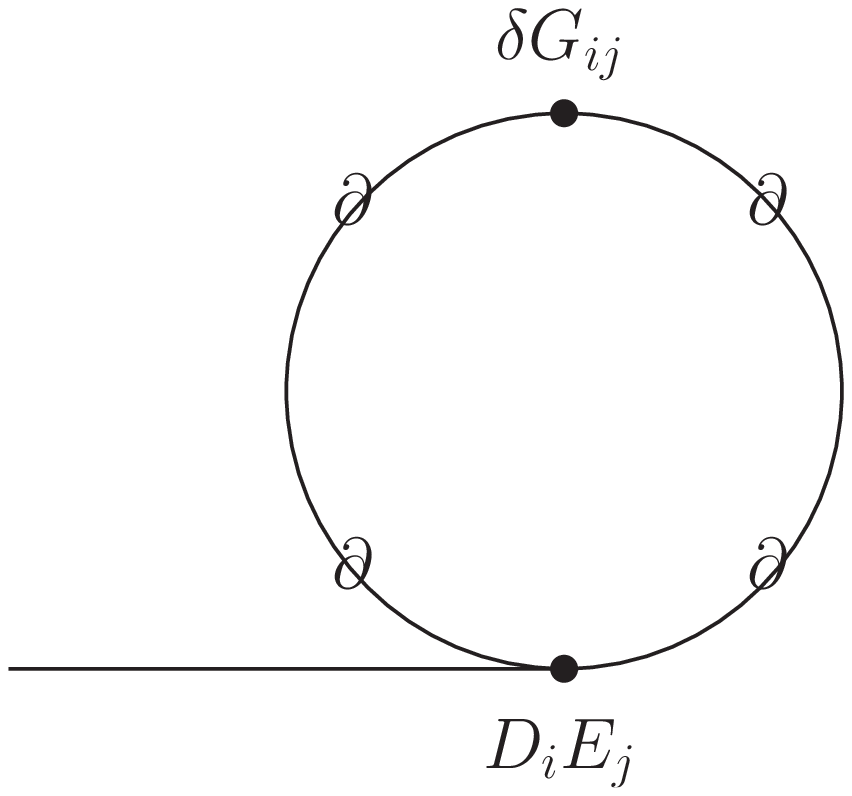}} 
\newlength{\boxbl}
\newsavebox{\boxbb}
\sbox{\boxbb}{\includegraphics[height=2.5cm]{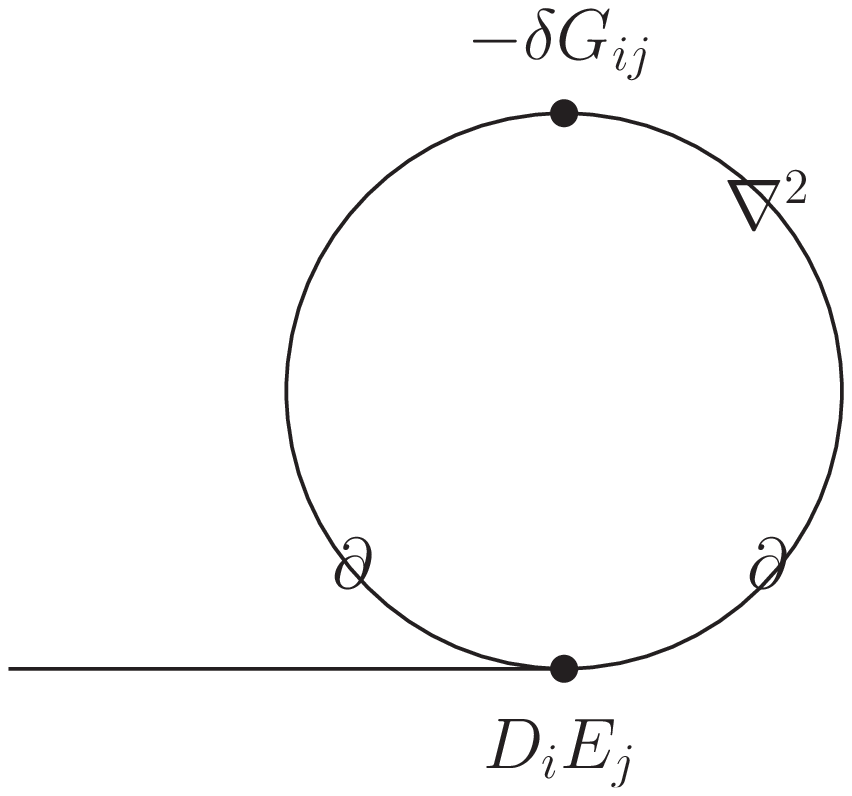}} 
\newlength{\boxbbl}
\newsavebox{\boxbbb}
\sbox{\boxbbb}{\includegraphics[height=2.5cm]{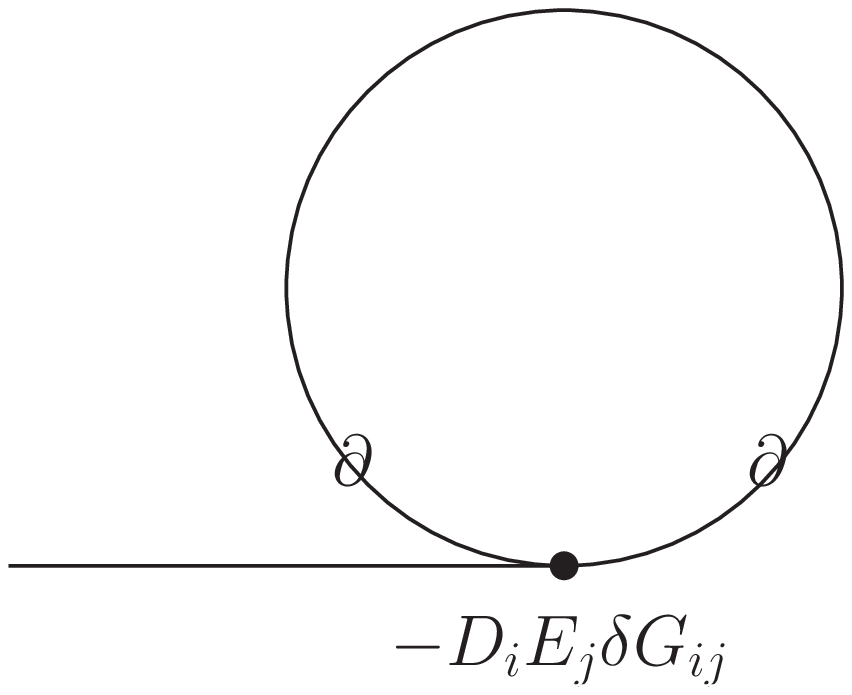}} 
\newlength{\boxbbbl}
\newsavebox{\boxc}
\sbox{\boxc}{\includegraphics[height=2.5cm]{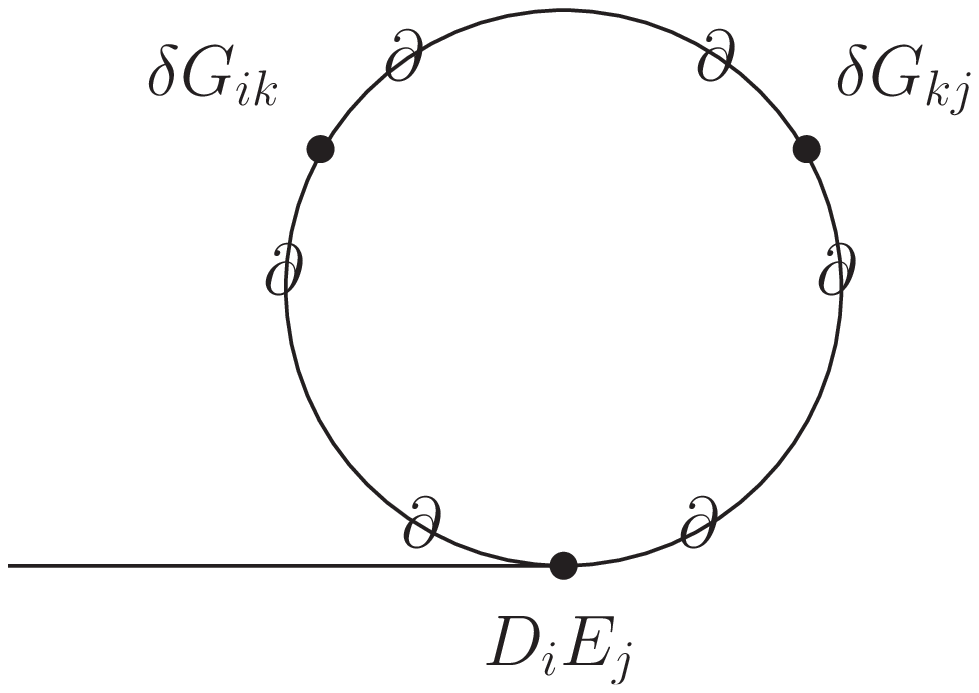}} 
\newlength{\boxcl}
\newsavebox{\boxcc}
\sbox{\boxcc}{\includegraphics[height=2.5cm]{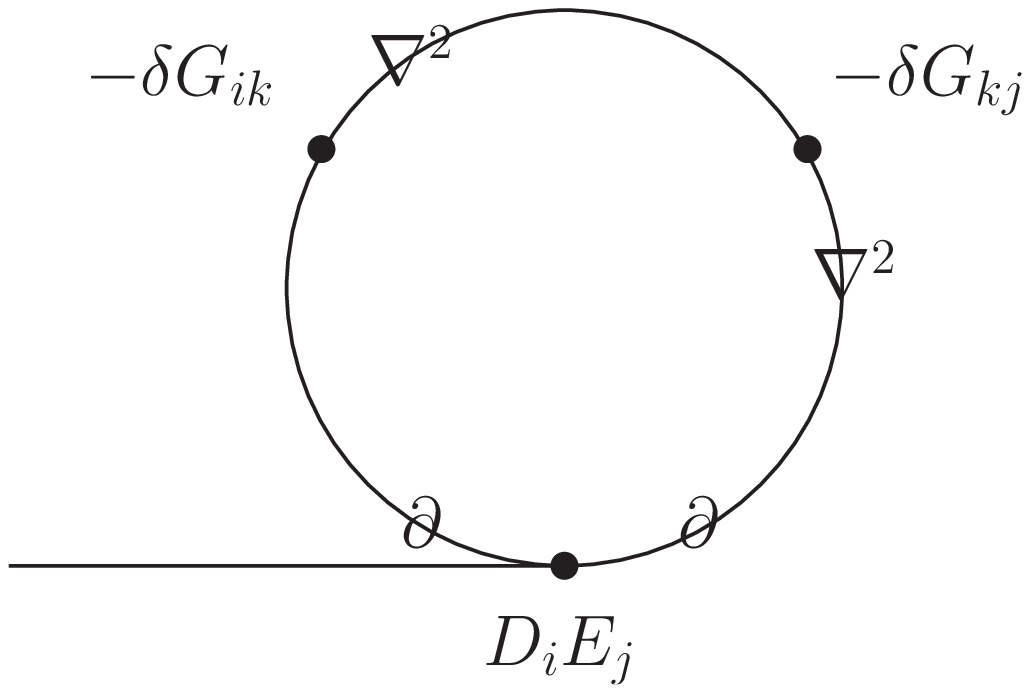}} 
\newlength{\boxccl}
\newsavebox{\boxccc}
\sbox{\boxccc}{\includegraphics[height=2.5cm]{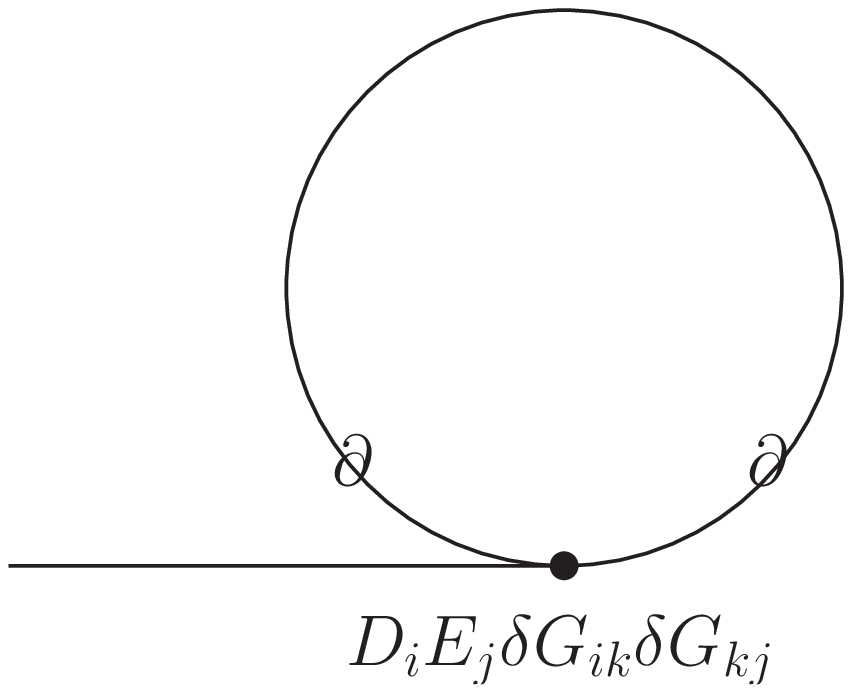}} 
\newlength{\boxcccl}
\newsavebox{\boxd}
\sbox{\boxd}{\includegraphics[height=2.5cm]{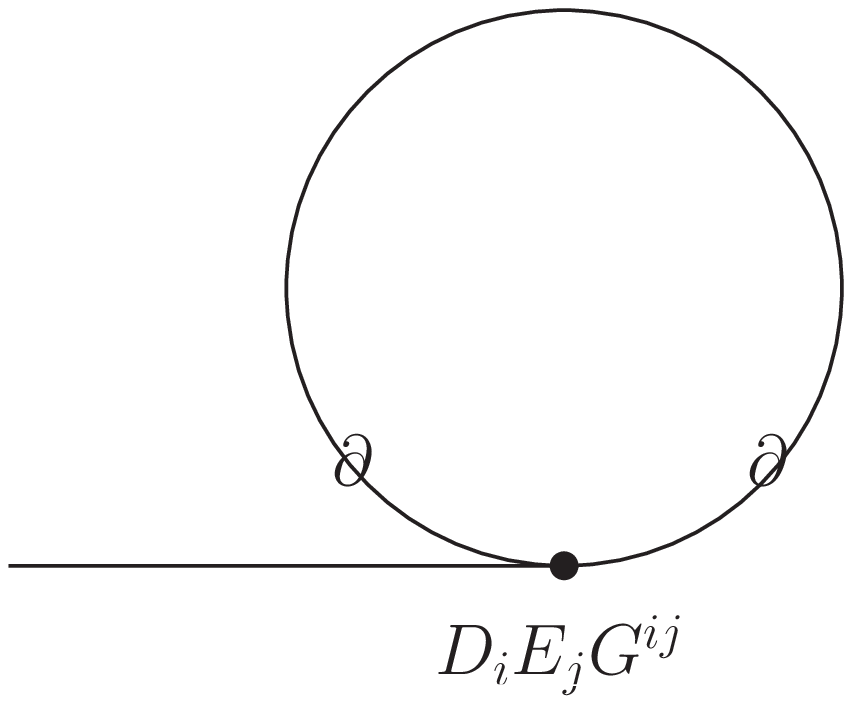}} 
\newlength{\boxdl}
\allowdisplaybreaks[3]

\begin{document}

\begin{titlepage}

\begin{flushright}
NCTS-TH/1814
\end{flushright}

\vspace{5em}

\begin{center}
{\Large \textbf{Infrared resummation for derivative interactions\\in de Sitter space}}
\end{center}

\begin{center}
Hiroyuki \textsc{Kitamoto}\footnote{E-mail: kitamoto@cts.nthu.edu.tw}, 
\end{center}

\begin{center}
\textit{Physics Division, National Center for Theoretical Sciences}\\
\textit{National Tsing-Hua University, Hsinchu 30013, Taiwan}\\
\end{center}

\begin{abstract}
In de Sitter space, scale invariant fluctuations give rise to infrared logarithmic corrections to physical quantities, 
which eventually spoil perturbation theories. 
For models without derivative interactions, it has been known that 
the field equation reduces to a Langevin equation with white noise in the leading logarithm approximation. 
The stochastic equation allows us to evaluate the infrared effects nonperturbatively. 
We extend the resummation formula so that it is applicable to models with derivative interactions. 
We first consider the nonlinear sigma model 
and next consider a more general model which consists of a noncanonical kinetic term and a potential term. 
The stochastic equations derived from the infrared resummation in these models 
can be understood as generalizations of the standard one to curved target spaces. 
\end{abstract}

\vspace{\fill}

July 2019

\end{titlepage}

\section{Introduction}
\setcounter{equation}{0}

In de Sitter (dS) space, the propagator for a massless and minimally coupled scalar field 
has a dS symmetry breaking term. 
This term is a direct consequence of the scale invariant spectrum at the superhorizon scale and is expressed 
as a logarithm of the scale factor of the Universe \cite{Vilenkin1982,Linde1982,Starobinsky1982}. 
In the presence of this scalar field, 
physical quantities may acquire time dependences through the propagator. 
In tribute to their origin, we call them quantum infrared (IR) effects in dS space. 

By employing the Schwinger-Keldysh formalism \cite{Schwinger1961,Keldysh1964}, 
we can study interacting field theories in dS space perturbatively.  
The IR effects at each loop level manifest as polynomials in the IR logarithm 
whose degrees increase with the loop level. 
At a late time, the leading IR effects come from the leading IR logarithms at each loop level. 
This fact indicates that the perturbative study breaks down 
after a large enough cosmic expansion. 
In order to understand such a situation, we need to evaluate the IR effects nonperturbatively. 

For models without derivative interactions, Starobinsky and Yokoyama proposed that 
the IR dynamics can be described nonperturbatively by a Langevin equation with white noise 
\cite{Starobinsky1986,Starobinsky1994}. 
Tsamis and Woodard proved that 
the stochastic approach is equivalent to the resummation of the leading IR logarithms to all-loop orders 
\cite{Woodard2005}. 

In this paper, we extend the resummation formula of the leading IR logarithms 
so that it is applicable to models with derivative interactions. 
As concrete examples, we consider the nonlinear sigma model 
and a more general model which consists of a noncanonical kinetic term and a potential term. 
Although there has been an attempt to derive stochastic equations in these models \cite{Tada2018},  
the consistency with the resummation of the leading IR logarithms has not been verified completely. 

We show that even when the kinetic term is noncanonical, 
the Yang-Feldman equation reduces to a Langevin equation with white noise 
in the leading logarithm approximation. 
The resulting stochastic equation is expressed in a covariant form 
with respect to the field coordinate transformation. 
As a specific feature of derivative interactions, we need to take into account 
the contribution from the subhorizon scale in evaluating the leading IR effects. 
This fact was found in \cite{Kitamoto2011,Kitamoto2012} 
where the energy-momentum tensor of the nonlinear sigma model was studied. 

This study takes a first step toward the understanding of the nonperturbative IR effects from gravity. 
The gravitational field theory includes massless and minimally coupled modes \cite{Woodard1994}, 
and it consists of derivative interactions in a similar way to the general scalar field theory. 
There are perturbative studies of gravitational IR effects; 
e.g. the gravitational IR corrections on the slow-roll parameters $\epsilon$ and $\eta$ 
were studied in single-field inflation models \cite{Kitamoto2017}. 
At the one-loop level, $\epsilon$ acquires a secular growth while $\eta$ does not. 
This result indicates that if $\epsilon$ and $\eta$ are vanishingly small at the beginning 
due to the pseudo shift symmetry, 
the quantum mechanism leads to an inflation model with a linear potential. 
In order to evaluate the whole time evolution of physical quantities, 
the IR resummation formula for gravity is necessary but it has not been known. 

The organization of this paper is as follows. 
In Sec. 2, we review the propagator for a massless and minimally coupled scalar field in dS space, 
in particular, its IR behavior. 
In Sec. 3, we derive a Langevin equation for the nonlinear sigma model  
by applying the leading logarithm approximation to its Yang-Feldman equation.  
In Sec. 4, we apply the same approach to a general scalar field theory 
which consists of a noncanonical kinetic term and a potential term. 
We also mention a relation to the Euclidean field theory on a sphere. 
We conclude with discussions in Sec. 5.  

\section{Free scalar field theory}
\setcounter{equation}{0}

Here we review a free scalar field which is massless and minimally coupled to the dS background.   
In particular, we focus on its IR dynamics which is a source of time dependent quantum effects.  
 
In the Poincar\'e coordinate, the metric of dS space is given by 
\begin{align}
ds^2=-dt^2+a^2(t)d\textbf{x}^2,\hspace{1em}a(t)=e^{Ht}, 
\label{metric}\end{align}
where the dimension of the spacetime is taken as an arbitrary $D$, 
and $H$ is the Hubble parameter which is constant in dS space. 
In the conformally flat coordinate, 
\begin{align}
g_{\mu\nu}=a^2(\tau)\eta_{\mu\nu},\hspace{1em}a(\tau)=\frac{1}{-H\tau}. 
\label{metric'}\end{align}
The conformal time $\tau$ is related to the cosmic time $t$ as $\tau=-\frac{1}{H}e^{-Ht}$. 
In this paper, $D$-dimensional vectors and tensors are expressed in the conformally flat coordinates. 

The quadratic action for a massless and minimally coupled scalar field is given by 
\begin{align}
S=\int \sqrt{-g}d^Dx\ \left[-\frac{1}{2}g^{\mu\nu}\partial_\mu\varphi\partial_\nu\varphi\right]. 
\end{align}
The free scalar field $\varphi_0$ can be expanded 
by the annihilation and the creation operators $a_\textbf{p},\ a_\textbf{p}^\dagger$ as 
\begin{align}
\varphi_0(x)=\int\frac{d^{D-1}p}{(2\pi)^{D-1}}\ 
\left[\phi_\textbf{p}(x)a_\textbf{p}+\phi_\textbf{p}^*(x)a_\textbf{p}^\dagger\right], 
\end{align}
\begin{align}
\phi_\textbf{p}(x)=\frac{\sqrt{\pi}}{2}H^\frac{D-2}{2}(-\tau)^\frac{D-1}{2}H_\frac{D-1}{2}^{(1)}(-p\tau)
e^{i\textbf{p}\cdot\textbf{x}}, 
\label{WF}\end{align}
where $p=|\textbf{p}|$ and $H_\frac{D-1}{2}^{(1)}$ is the Hankel function of the first kind. 
See, e.g., \cite{ToI} for details of special functions. 
Throughout this paper, we consider the Bunch-Davies vacuum $a_\textbf{p}|0\rangle=0$. 
It should be noted that $p$ denotes the comoving momentum, 
and the physical momentum is given by $P=p/a(\tau)$. 

In this paper, we mainly consider the contribution from the superhorizon scale 
where curved spacetime-specific effects become manifest. 
At the superhorizon scale $P\ll H \Leftrightarrow -p\tau\ll 1$, the wave function behaves as 
\begin{align}
\phi_\textbf{p}(x)\simeq -i\frac{2^\frac{D-3}{2}\Gamma(\frac{D-1}{2})}{\sqrt{\pi}}
\frac{H^\frac{D-2}{2}}{p^\frac{D-1}{2}}e^{i\textbf{p}\cdot\textbf{x}}. 
\label{WF-IR}\end{align}
The quantum fluctuation of a massless and minimally coupled scalar field is frozen at the superhorizon scale 
as the leading IR term of the wave function has no time dependence.  

The contribution from (\ref{WF-IR}) to the propagator at the coincident point is given by  
\begin{align}
\langle\varphi_0^2(x)\rangle\simeq \int_{p<Ha(\tau)}\frac{d^{D-1}p}{(2\pi)^{D-1}}\ 
\frac{2^{D-3}\Gamma^2(\frac{D-1}{2})}{\pi}\frac{H^{D-2}}{p^{D-1}}. 
\label{P-IR'}\end{align}
The upper bound of the momentum integral is taken on the horizon 
because the $1/p^{D-1}$ spectrum is dominant at the superhorizon scale. 
The propagator has an IR logarithmic divergence due to the presence of the scale invariant spectrum. 

In order to regularize the IR divergence, 
we introduce an IR cutoff $p_0$ which fixes the minimum value of the comoving momentum. 
Under the current setting, the IR contribution to the propagator is expressed 
as the logarithm of the scale factor \cite{Vilenkin1982,Linde1982,Starobinsky1982}: 
\begin{align}
\langle\varphi_0^2(x)\rangle
&\simeq \frac{2H^{D-2}}{(4\pi)^\frac{D}{2}}\frac{\Gamma(D-1)}{\Gamma(\frac{D}{2})}
\int^{Ha(\tau)}_{p_0}\frac{dp}{p} \notag\\
&=\frac{2H^{D-2}}{(4\pi)^\frac{D}{2}}\frac{\Gamma(D-1)}{\Gamma(\frac{D}{2})}
\int^H_{p_0/a(\tau)}\frac{dP}{P} \notag\\
&=\frac{2H^{D-2}}{(4\pi)^\frac{D}{2}}\frac{\Gamma(D-1)}{\Gamma(\frac{D}{2})}\log (a(\tau)/a_0). 
\label{P-IR}\end{align}
From (\ref{P-IR'}) to (\ref{P-IR}), we used 
\begin{align}
\Gamma(2z)=\frac{2^{2z-1}}{\sqrt{\pi}}\Gamma(z)\Gamma(z+\frac{1}{2}). 
\label{formula}\end{align}
This formula of the gamma function is used repeatedly in this paper. 
The scale factor at the initial time $a_0$ is related to the IR cutoff as $a_0=p_0/H$. 
Physically, $a(\tau)/p_0$ means the size of the Universe, and it expands with the scale factor. 
Since the Hubble scale $1/H$ is constant, the degrees of freedom at the superhorizon scale increase with time. 
This increase gives rise to the logarithmic dependence of the scale factor to the propagator 
through the scale invariant spectrum. 

Here we write down the explicit form of the propagator 
for a massless and minimally coupled scalar field \cite{Woodard2010}: 
\begin{align}
\langle\varphi_0(x)\varphi_0(x')\rangle
= I(y)+\frac{H^{D-2}}{(4\pi)^\frac{D}{2}}\frac{\Gamma(D-1)}{\Gamma(\frac{D}{2})}\log (a(\tau)a(\tau')/a_0^2), 
\label{propagator}\end{align} 
\begin{align}
I(y)&=\frac{H^{D-2}}{(4\pi)^\frac{D}{2}}\left\{\Gamma(\frac{D}{2}-1)\left(\frac{y}{4}\right)^{1-\frac{D}{2}}
-\frac{\Gamma(\frac{D}{2}+1)}{2-\frac{D}{2}}\left(\frac{y}{4}\right)^{2-\frac{D}{2}}
+\frac{\Gamma(D-1)}{\Gamma(\frac{D}{2})}\delta\right. \\
&\hspace{5em}\left.+\sum^{\infty}_{n=1}
\left[\frac{\Gamma(D-1+n)}{n\Gamma(\frac{D}{2}+n)}\left(\frac{y}{4}\right)^n
-\frac{\Gamma(\frac{D}{2}+1+n)}{(2-\frac{D}{2}+n)(n+1)!}\left(\frac{y}{4}\right)^{2-\frac{D}{2}+n}
\right]\right\}, \notag
\end{align}
\begin{align}
\delta\equiv-\psi(1-\frac{D}{2})+\psi(\frac{D-1}{2})+\psi(D-1)+\psi(1),\hspace{1em}
\psi(z)\equiv \frac{1}{\Gamma(z)}\frac{d\Gamma(z)}{dz}, 
\end{align} 
where $y$ denotes the square of the physical distance: 
\begin{align}
y=H^2a(\tau)a(\tau')\Delta x^2,\hspace{1em}\Delta x^2\equiv -(\tau-\tau')^2+(\textbf{x}-\textbf{x}')^2. 
\end{align}
The metric of dS space and $y$ are invariant under the rescaling $\tau\to C\tau,\ \textbf{x}\to C\textbf{x}$. 
The propagator for a massless and minimally coupled scalar field (\ref{propagator}) does not respect 
this dS symmetry. 
This is due to the IR cutoff dependence of the second term. 

We evaluate the twice-differentiated propagator at the coincident point 
$\langle\partial_\mu\varphi_0(x)\partial_\nu\varphi_0(x)\rangle$ 
as we use it in the subsequent discussion. 
In the dimensional regularization, the $y^\alpha$ term disappears at the coincident point 
if $\alpha$ becomes positive for a sufficiently low $D$. 
Considering this fact, the twice-differentiated propagator at the coincident point is evaluated as 
\begin{align}
\langle\partial_\mu\varphi_0(x)\partial_\nu\varphi_0(x)\rangle
&=\frac{H^{D-2}}{(4\pi)^\frac{D}{2}}\frac{\Gamma(D)}{4\Gamma(\frac{D}{2}+1)}
\partial_\mu\partial'_\nu y^1|_{x'\to x} \notag\\
&=-\frac{H^D}{(4\pi)^\frac{D}{2}}\frac{\Gamma(D)}{D\Gamma(\frac{D}{2})}g_{\mu\nu}. 
\label{tdp}\end{align}
The twice-differentiated propagator respects the dS symmetry. 
That is because if differential operators act on both $x$ and $x'$, 
the dS symmetry breaking term in (\ref{propagator}) does not have any contribution, 
\begin{align}
\partial_\mu\partial'_\nu \log (a(\tau)a(\tau')/a_0^2)=0. 
\label{disappear}\end{align}

In interacting field theories with massless and minimally coupled scalar fields, 
physical quantities may acquire IR logarithmic corrections through the propagator.  
At the late time $\log(a(\tau)/a_0)\gg 1$, the leading IR effects come from the leading IR logarithms 
at each loop level. 
As an example, let us consider the $\lambda \varphi^i\varphi^i\varphi^j\varphi^j$ interaction.  
With each increase in the loop level, 
quantum corrections from the nonderivative interaction are multiplied by at most a factor of 
\begin{align}
\lambda H^{D-4} \log^2(a(\tau)/a_0). 
\end{align} 
As another example, 
let us consider the $f^2\varphi^i\varphi^ig^{\mu\nu}\partial_\mu\varphi^j\partial_\nu\varphi^j$ interaction. 
With each increase in the loop level, 
quantum corrections from the derivative interaction are multiplied by at most a factor of 
\begin{align}
f^2H^{D-2} \log (a(\tau)/a_0). 
\end{align}
These secular growths indicate that 
even though the dimensionless couplings are small as $\lambda H^{D-4},\ f^2H^{D-2}\ll 1$, 
the perturbative study breaks down after a long enough time: 
\begin{align}
\lambda H^{D-4} \log^2(a(\tau)/a_0) \sim 1\hspace{1em}\text{or}\hspace{1em}
f^2H^{D-2} \log (a(\tau)/a_0) \sim 1.  
\end{align}
A rational approach for evaluating such nonperturbative effects is 
to resum the leading IR logarithms to all-loop orders. 
The resummation formula for nonderivative interactions 
has been derived by Tsamis and Woodard \cite{Woodard2005}. 
In subsequent sections, we derive the resummation formula for derivative interactions. 

\section{Nonlinear sigma model}
\setcounter{equation}{0}

As a concrete model with derivative interactions, we consider the nonlinear sigma model: 
\begin{align}
S=\int \sqrt{-g}d^Dx\ 
\left[-\frac{1}{2}G_{ij}(\varphi)g^{\mu\nu}\partial_\mu\varphi^i\partial_\nu\varphi^j\right], 
\label{NLsigma}\end{align}
where $G_{ij}\ (i=1,\cdots,N)$ is the metric of the target space. 
The action is invariant under the general coordinate transformation 
where $\varphi^i$ are identified as coordinates. 
The global symmetry guarantees that this model consists of massless and minimally coupled scalar fields 
without fine-tuning the quadratic action. 
For later use, we introduce the vielbein: 
\begin{align}
G_{ij}(\varphi)=E_i^{\ a}(\varphi)E_j^{\ b}(\varphi)\delta_{ab}, 
\end{align}
where $a$ denotes the index of the flat tangent space. 

The field equation for the nonlinear sigma model is given by 
\begin{align}
\frac{1}{\sqrt{-g}}\partial_\mu\left\{\sqrt{-g}\ G_{ij}(\varphi)g^{\mu\nu}\partial_\nu\varphi^j\right\}
-\frac{1}{2}\frac{\partial}{\partial\varphi^i}G_{jk}(\varphi)g^{\mu\nu}\partial_\mu\varphi^j\partial_\nu\varphi^k=0. 
\label{FE0}\end{align}
As is well known, the equation can be rewritten in the following form:  
\begin{align}
\nabla^\mu\nabla_\mu\varphi^i
+\Gamma^i_{\ jk}(\varphi)g^{\mu\nu}\partial_\mu\varphi^j\partial_\nu\varphi^k=0,  
\label{FE0'}\end{align}
where $\nabla_\mu$ is the covariant derivative in the spacetime 
and $\Gamma^i_{\ jk}$ is the Levi-Civita connection in the target space. 
However, in order to find a covariant Yang-Feldman equation, it is more useful to rewrite (\ref{FE0}) as 
\begin{align}
\frac{1}{\sqrt{-g}}\partial_\mu\big\{\sqrt{-g}\ E_i^{\ a}(\varphi)g^{\mu\nu}\partial_\nu\varphi^i\big\}-D_iE_j^{\ a}(\varphi)g^{\mu\nu}\partial_\mu\varphi^i\partial_\nu\varphi^j=0, 
\label{FE1}\end{align}
where $D_i$ is the covariant derivative in the target space. 
The covariance is manifest in (\ref{FE1}) in contrast to (\ref{FE0}) and (\ref{FE0'}). 

The Yang-Feldman equation for the nonlinear sigma model is given by 
\begin{align}
-i\int d^D&x'\ \partial'_\mu\left\{\sqrt{-g(x')}g^{\mu\nu}(x')\partial'_\nu G^R(x,x') E_i^{\ a}(\varphi(x'))\right\}
\varphi^i(x') \notag\\
&=\varphi_0^a(x)
-i\int\sqrt{-g(x')}d^Dx'\ G^R(x,x')D_iE_j^{\ a}(\varphi(x'))
g^{\mu\nu}(x')\partial'_\mu\varphi^i(x')\partial'_\nu\varphi^j(x'), 
\label{YF1}\end{align}
where $G^R$ denotes the retarded propagator: 
\begin{align}
G^R(x,x')=\theta(\tau-\tau')\langle[\varphi_0(x),\varphi_0(x')]\rangle, 
\label{retarded}\end{align}
which satisfies 
\begin{align}
\nabla^\mu\nabla_\mu G^R(x,x')=\frac{i\delta^{(D)}(x-x')}{\sqrt{-g(x)}}. 
\label{kernel}\end{align}
We can reproduce (\ref{FE1}) by applying the d'Alembert operator to both sides of (\ref{YF1}).  
It should be noted that the homogeneous solution, i.e. the free field, has the index $a$ rather than $i$, 
and both sides of (\ref{YF1}) have the index $a$. 

The Yang-Feldman equation describes quantum effects from all momentum scales, 
and it is not exactly solvable in general. 
We show that in the leading logarithm approximation, the Yang-Feldman equation 
reduces to a Langevin equation with white noise, 
which is a more suitable tool for studying the IR dynamics. 

For the first term in the right-hand side of (\ref{YF1}), we extract its leading IR behavior as 
\begin{align}
\varphi_0^a(x)\simeq\bar{\varphi}_0^a(x) 
\equiv\int\frac{d^{D-1}p}{(2\pi)^{D-1}}\ \theta(Ha(\tau)-p)
\left[-i\frac{2^\frac{D-3}{2}\Gamma(\frac{D-1}{2})}{\sqrt{\pi}}\frac{H^\frac{D-2}{2}}{p^\frac{D-1}{2}}
e^{i\textbf{p}\cdot\textbf{x}}a_\textbf{p}^a+\text{(H.c.)}\right], 
\label{right1}\end{align}
where $[a_\textbf{p}^a,a_{\textbf{p}'}^{\dagger b}]
=(2\pi)^{D-1}\delta^{(D-1)}(\textbf{p}-\textbf{p}')\delta^{ab}$ and the other commutation relations are zero. 
Since the $1/p^\frac{D-1}{2}$ term is dominant at the superhorizon scale, 
we introduce the step function which is nonzero only at this IR scale. 
Substituting (\ref{right1}), 
each Wightman function without differential operators $\langle\varphi_0(x)\varphi_0(x')\rangle$ 
provides a single IR logarithm.  

The second term in the right-hand side of (\ref{YF1}) includes two differentiated fields
$\partial'_\mu\varphi^i(x')\partial'_\nu\varphi^j(x')$. 
At each order of the expansion in the interaction vertex 
\begin{align}
\frac{1}{2}\delta G_{ij}(\varphi)g^{\mu\nu}\partial_\mu\varphi^i\partial_\nu\varphi^j,\hspace{1em} 
\delta G_{ij}(\varphi)=G_{ij}(\varphi)-\delta_{ij},  
\label{vertex}\end{align}
the diagram with the leading IR logarithms includes a loop of twice-differentiated propagators, 
which starts at $x'$ and ends at $x'$: 
\begin{align}
\parbox{\boxal}{\usebox{\boxa}}\ ,\hspace{1em}
\parbox{\boxbl}{\usebox{\boxb}}\ ,\hspace{1em}
\parbox{\boxcl}{\usebox{\boxc}}\ ,\hspace{1em}\cdots\hspace{1em}. 
\label{diagrams}\end{align}
Here the horizontal line segment denotes $G^R(x,x')$, and $\partial$ denotes a differential operator. 
The other diagrams, 
which are obtained by transferring any of the differential operators from the propagators inside the loop 
to the nondifferentiated fields outside the loop, 
have reduced powers of the IR logarithms.
In other words, we need to minimize the number of differentiated propagators 
to obtain the leading IR logarithms.\footnote{
The statement holds true only if the retarded propagator starting at $x'$ is not differentiated. 
Therefore, it does not apply to the left-hand side of (\ref{YF1}) with $\partial'_\nu G^R(x,x')$. }

Using partial integrations, we can rewrite the diagrams as 
\begin{align}
\parbox{\boxbl}{\usebox{\boxb}}\ 
\simeq\parbox{\boxbbl}{\usebox{\boxbb}}\ 
\simeq\parbox{\boxbbbl}{\usebox{\boxbbb}}\ , 
\end{align}
\begin{align}
\parbox{\boxcl}{\usebox{\boxc}}\ 
\simeq\parbox{\boxccl}{\usebox{\boxcc}}\ 
\simeq\parbox{\boxcccl}{\usebox{\boxccc}}\ , 
\end{align}
where we used the fact that the diagrams with differentiated $\delta G_{ij}$ are negligible 
in the leading logarithm approximation. 
The presence of the propagator with the d'Alembert operator $\nabla^2$ allows us 
to evaluate the vertex integral trivially.\footnote{
The calculation technique can also be adopted in evaluating the energy-momentum tensor 
of the nonlinear sigma model \cite{Kitamoto2012}. } 
Applying the same procedure also at higher orders of the expansion in the interaction vertex,  
the total contribution of the diagrams (\ref{diagrams}) can be summarized as 
\begin{align}
\parbox{\boxdl}{\usebox{\boxd}}\ , 
\label{diagram}\end{align} 
where we used 
\begin{align}
\delta_{ij}-\delta G_{ij}+\delta G_{ik}\delta G_{kj}+\cdots=G^{ij}. 
\end{align}

In the second term in the right-hand side of (\ref{YF1}), 
the diagrammatic discussion can be expressed as   
\begin{align}
\partial'_\mu\varphi^i(x')\partial'_\nu\varphi^j(x')
&\simeq G^{ij}(\varphi(x'))\langle\partial'_\mu\varphi_0(x')\partial'_\nu\varphi_0(x')\rangle \notag\\
&=-G^{ij}(\varphi(x'))\frac{H^D}{(4\pi)^\frac{D}{2}}\frac{\Gamma(D)}{D\Gamma(\frac{D}{2})}g_{\mu\nu}(x'). 
\label{right2}\end{align}
We substituted the value of the twice-differentiated propagator (\ref{tdp}) in the last line. 
It should be noted that (\ref{right2}) is the approximation 
for the diagrams with a loop of twice-differentiated propagators (\ref{diagrams}). 
This approximation means that
we can integrate out differentiated fields as if nondifferentiated fields were constant. 
In other words, we cannot treat $G_{ij}$ as a constant coefficient of the kinetic term 
in evaluating correlation functions of nondifferentiated fields. 

The second term in the right-hand side of (\ref{YF1}) also includes the retarded propagator. 
In order to evaluate it, we need to know the real and the imaginary parts of the wave function. 
Therefore, the wave function (\ref{WF}) should be expanded further than in (\ref{WF-IR}):  
\begin{align}
\phi_\textbf{p}(x)\simeq 
-i\left\{\frac{2^\frac{D-3}{2}\Gamma(\frac{D-1}{2})}{\sqrt{\pi}}\frac{H^\frac{D-2}{2}}{p^\frac{D-1}{2}}
+i\frac{\sqrt{\pi}}{2^\frac{D+1}{2}\Gamma(\frac{D+1}{2})}H^\frac{D-2}{2}p^\frac{D-1}{2}(-\tau)^{D-1}\right\}
e^{i\textbf{p}\cdot\textbf{x}}. 
\label{WF-IR'}\end{align}
Substituting it in (\ref{retarded}), the retarded propagator is evaluated as 
\begin{align}
G^R(x,x')&\simeq \theta(\tau-\tau')\int\frac{d^{D-1}p}{(2\pi)^{D-1}}\ 
\frac{-iH^{D-2}}{D-1}\left[(-\tau')^{D-1}-(-\tau)^{D-1}\right]e^{i\textbf{p}\cdot(\textbf{x}-\textbf{x}')} \notag\\
&=\frac{-i}{(D-1)H}\theta(\tau-\tau')\left[a^{-(D-1)}(\tau')-a^{-(D-1)}(\tau)\right]
\delta^{(D-1)}(\textbf{x}-\textbf{x}'). 
\label{right3'}\end{align}
The retarded propagator itself does not have an IR logarithm, in contrast to the Wightman function, 
while the vertex integral with it induces a single IR logarithm as 
\begin{align}
\int\sqrt{-g(x')}d^Dx'\ G^R(x,x')
&\simeq -\frac{i}{(D-1)H}\int^\tau a(\tau')d\tau'\ 
\left[1-\left(a(\tau')/a(\tau)\right)^{D-1}\right] \notag\\
&\simeq-\frac{i}{(D-1)H}\int^tdt', 
\label{right3}\end{align}
where we dropped the trivial spatial integral $\int d^{D-1}x'\ \delta^{(D-1)}(\textbf{x}-\textbf{x}')$.  
In the last line, we neglected the second term because it does not induce an IR logarithm. 
As $dt'=d(\log a(t'))/H$, the vertex integral provides a single IR logarithm 
in addition to the contribution from the integrand 
$D_iE_j^{\ a}(\varphi(x'))g^{\mu\nu}(x')\partial'_\mu\varphi^i(x')\partial'_\nu\varphi^j(x')$. 
It should be noted that the higher-order terms neglected in (\ref{right3'}) reduce 
the powers of the IR logarithms from the integrand.  

The left-hand side of (\ref{YF1}) can be rewritten as 
\begin{align}
E_i^{\ a}(\varphi(x))\varphi^i(x)
-i\int \sqrt{-g(x')}d^Dx'\ g^{\mu\nu}(x')\partial'_\mu G^R(x,x') \partial'_\nu E_i^{\ a}(\varphi(x'))\varphi^i(x'). 
\label{YF1'}\end{align}
We can extract the leading IR logarithms from the first term just by iteratively substituting $\varphi$ in it. 
In order to evaluate the leading IR effects from the second term, 
we approximate the differentiated retarded propagator as 
\begin{align}
\partial'_\mu G^R(x,x')
&\simeq \theta(\tau-\tau')\delta_\mu^{\ 0}\int\frac{d^{D-1}p}{(2\pi)^{D-1}}\ 
iH^{D-2}(-\tau')^{D-2}e^{i\textbf{p}\cdot(\textbf{x}-\textbf{x}')} \notag\\
&= i\theta(\tau-\tau')\delta_\mu^{\ 0} 
a^{-(D-2)}(\tau')\delta^{(D-1)}(\textbf{x}-\textbf{x}'). 
\label{left'}\end{align}
As we did in (\ref{WF-IR'}) and (\ref{right3'}), we expanded the differentiated retarded propagator 
in powers of $(-p\tau)$ and $(-p\tau')$ and kept the zeroth-order term.\footnote{
Each contribution from the $(-p\tau')^n$ term for $n\ge 1$ induces the leading IR logarithms, 
while the leading IR logarithms from the higher-order terms are canceled in total.
The cancellation holds true in scalar field theories where the Lorentz indices are contracted 
within differential operators.}
Substituting (\ref{left'}), the vertex integral with the differentiated retarded propagator is given by 
\begin{align}
\int \sqrt{-g(x')}d^Dx'\ g^{\mu\nu}(x')\partial'_\mu G^R(x,x')
&\simeq -i\int^\tau d\tau'\ \delta_0^{\ \nu}\notag\\
&=-i\int^tdt'\ a^{-1}(t')\delta_0^{\ \nu}, 
\label{left}\end{align}
where we dropped the trivial spatial integral. 
For the integrand $\partial'_0 E_i^{\ a}(\varphi(x'))\varphi^i(x')$, 
the differential operator $\partial'_0=a(t')\partial'_t$ removes a single IR logarithm from $E_i^{\ a}(\varphi(x'))$ 
and adds one scale factor $a(t')$. 
The scale factor from the differential operator is canceled by $a^{-1}(t')$ in (\ref{left}), 
and then the vertex integral provides a single IR logarithm as $dt'=d(\log a(t'))/H$.  
That is how the second term in (\ref{YF1'}) induces the leading IR logarithms. 

Applying the leading logarithm approximation (\ref{right1}), (\ref{right2}), (\ref{right3}) and (\ref{left}), 
the Yang-Feldman equation (\ref{YF1}) reduces to 
\begin{align}
E_i^{\ a}(\varphi(x))\varphi^i&(x)
-\int^tdt'\ \partial'_tE_i^{\ a}(\varphi(t',\textbf{x}))\varphi^i(t',\textbf{x}) \notag\\
&=\bar{\varphi}_0^a(x)
+\frac{H^{D-1}}{(4\pi)^\frac{D}{2}}\frac{\Gamma(D-1)}{\Gamma(\frac{D}{2})}
\int^tdt'\ D_iE^{ia}(\varphi(t',\textbf{x})). 
\end{align}
Differentiating both sides with respect to $t$, we obtain the following equation: 
\begin{align}
E_i^{\ a}(\varphi(x))\dot{\varphi}^i(x)=\dot{\bar{\varphi}}_0^a(x)
+\frac{H^{D-1}}{(4\pi)^\frac{D}{2}}\frac{\Gamma(D-1)}{\Gamma(\frac{D}{2})}D_iE^{ia}(\varphi(x)), 
\label{Langevin1}\end{align}
where $\dot{\varphi}^i=\partial_t \varphi^i$ 
and the correlation function of $\dot{\bar{\varphi}}_0^a$ is given by 
\begin{align}
\langle\dot{\bar{\varphi}}_0^a(t,\textbf{x})\dot{\bar{\varphi}}_0^b(t',\textbf{x})\rangle
&=\delta^{ab}\int \frac{d^{D-1}p}{(2\pi)^{D-1}}\ (H^2a(t))^2\delta(Ha(t)-p)\delta(Ha(t)-Ha(t')) \notag\\
&\hspace{8em}\times \frac{2^{D-3}\Gamma^2(\frac{D-1}{2})}{\pi}\frac{H^{D-2}}{p^{D-1}}\notag\\
&=\frac{2H^{D-1}}{(4\pi)^\frac{D}{2}}\frac{\Gamma(D-1)}{\Gamma(\frac{D}{2})}\delta^{ab}\delta(t-t'). 
\label{white1}\end{align} 
It should be recalled that the time dependence of $\bar{\varphi}_0^a$ appears only through the step function 
which is nonzero only at the superhorizon scale. 
As a consequence of this fact, 
the correlation function of $\dot{\bar{\varphi}}_0^a$ is proportional to the temporal delta function. 
This type of fluctuation is called a white noise. 
The equation in (\ref{Langevin1}) and (\ref{white1}) is known 
as a Langevin equation with white noise \cite{Risken}. 

From the Langevin equation, 
we can derive the equation satisfied by the probability density $\rho$: 
\begin{align}
\dot{\rho}(t,\phi)
=\ &\frac{H^{D-1}}{(4\pi)^\frac{D}{2}}\frac{\Gamma(D-1)}{\Gamma(\frac{D}{2})}
\frac{\partial^2}{\partial\phi^i\partial\phi^j}\left\{G^{ij}(\phi)\rho(t,\phi)\right\} \notag\\
&-\frac{H^{D-1}}{(4\pi)^\frac{D}{2}}\frac{\Gamma(D-1)}{\Gamma(\frac{D}{2})}
\frac{\partial}{\partial\phi^i}\left\{\left[\frac{\partial}{\partial \phi^j}E^i_{\ a}(\phi) E^{ja}(\phi)
+E^i_{\ a}(\phi)D_jE^{ja}(\phi)\right]\rho(t,\phi)\right\} \notag\\
=\ &\frac{H^{D-1}}{(4\pi)^\frac{D}{2}}\frac{\Gamma(D-1)}{\Gamma(\frac{D}{2})}
\frac{\partial^2}{\partial\phi^i\partial\phi^j}\left\{G^{ij}(\phi)\rho(t,\phi)\right\} \notag\\
&+\frac{H^{D-1}}{(4\pi)^\frac{D}{2}}\frac{\Gamma(D-1)}{\Gamma(\frac{D}{2})}
\frac{\partial}{\partial\phi^i}\left\{\Gamma^i_{\ jk}(\phi)G^{jk}(\phi)\rho(t,\phi)\right\}, 
\label{FP1}\end{align}
which is known as a Fokker-Planck equation. 
By using its solution, we can evaluate the vacuum expectation value (vev) of any operator 
made of $\varphi^i$ as 
\begin{align}
\langle F^{i_1\cdots i_n}(\varphi(x))\rangle=\int d^N\phi\ F^{i_1\cdots i_n}(\phi)\rho(t,\phi), 
\label{F}\end{align}
where $F^{i_1\cdots i_n}$ denotes an arbitrary function. 
It should be noted that $\varphi^i$ are operators while $\phi^i$ are $c$-numbers. 
For a general case, 
we review the relation between the Langevin equation and the Fokker-Planck equation in the Appendix. 
We can rewrite (\ref{FP1}) in the following covariant form: 
\begin{align}
\frac{\dot{\rho}(t,\phi)}{\sqrt{G(\phi)}}
=\frac{H^{D-1}}{(4\pi)^\frac{D}{2}}\frac{\Gamma(D-1)}{\Gamma(\frac{D}{2})}
D^iD_i\left\{\frac{\rho(t,\phi)}{\sqrt{G(\phi)}}\right\}. 
\label{FP1'}\end{align}
It should be noted that as seen in (\ref{F}), $\rho$ itself is not a scalar quantity 
but $\rho/\sqrt{G}$ is a scalar quantity. 

The Fokker-Planck equation is exactly solvable for an equilibrium state 
which is eventually reached: $\rho(t,\phi)\to \rho_\infty(\phi)$ at $t\to\infty$ if it exists. 
Since $-D^iD_i=D^{\dagger i} D_i$ is non-negative--definite, the solution for the equilibrium state is given by 
\begin{align}
D_i\left\{\frac{\rho_\infty(t,\phi)}{\sqrt{G(\phi)}}\right\}=0\hspace{1em}
\Rightarrow\hspace{1em} \rho_\infty(\phi)=Z^{-1}\sqrt{G(\phi)},  
\label{solution1}\end{align}
where $Z$ is constant. 
Since the total integral of the probability density is kept at unity, the overall coefficient is fixed as 
\begin{align}
Z=\int d^N\phi\ \sqrt{G(\phi)}. 
\label{solution1'}\end{align}

As seen in (\ref{solution1}) and (\ref{solution1'}), 
the convergence of $\sqrt{G(\phi)}$ at $|\phi|\to\infty$ determines 
whether an equilibrium state is eventually reached or not. 
If the asymptotic behavior of $\sqrt{G(\phi)}$ is at most $|\phi|^{-\beta}$, $\beta>0$, 
no equilibrium state is reached. 

Here we mention the previous study of the stochastic equation for the same model \cite{Tada2018}. 
The previous study did not take into account the contribution from the second term 
in the left-hand side of (\ref{FE1}). 
As discussed in (\ref{vertex})--(\ref{right2}), 
this term also induces the leading IR logarithms and leads to the drift term in (\ref{Langevin1}). 
In the previous study, this drift term was introduced by hand, 
backwards from the Fokker-Planck equation (\ref{FP1'}). 
Let us recall that the diagrams with a loop of twice-differentiated propagators (\ref{diagrams}) 
induce this drift term. 
As seen in (\ref{disappear}), the IR logarithm does not have any contribution 
to the twice-differentiated propagator. 
This fact shows that we need to take into account the contribution from the subhorizon scale 
if derivative interactions are present.  

We give a concrete example which induces an equilibrium state: 
\begin{align}
G_{ij}(\varphi)=\exp\left(-\frac{f^2}{2N}\varphi^k\varphi^k\right)\delta_{ij}, 
\label{compact}\end{align}
where $f$ is a coupling constant.  
In this case, the saturation values of $\langle (\varphi^i(x)\varphi^i(x))^n \rangle$ are given by 
\begin{align}
\rho_\infty(\phi)=Z^{-1}\exp\left(-\frac{f^2}{4}\phi^k\phi^k\right)
\hspace{1em}\Rightarrow\hspace{1em}
\langle (\varphi^i(x)\varphi^i(x))^n \rangle|_{t\to\infty}
=\frac{\Gamma(\frac{N}{2}+n)}{\Gamma(\frac{N}{2})}\left(\frac{2}{f}\right)^{2n}. 
\label{example}\end{align}

The $1/f^{2n}$ dependences of (\ref{example}) can be understood in the following way. 
Up to the leading logarithm accuracy, 
the time evolution of $\langle (\varphi^i(x)\varphi^i(x))^n \rangle$ is given by 
\begin{align}
\langle (\varphi^i(x)\varphi^i(x))^n \rangle
\simeq H^{n(D-2)}\sum_{m=0}^\infty c_m (f^2H^{D-2})^m\log^{n+m}(a(t)/a_0), 
\label{estimate1}\end{align}
where $c_m$ denote numerical coefficients. 
In the case that an equilibrium state is eventually reached, 
the vev grows until the time when the perturbation theory breaks down: 
\begin{align}
f^2H^{D-2}\log (a(t_c)/a_0)\sim 1\hspace{1em}\Leftrightarrow\hspace{1em}
t_c-t_0\sim \frac{1}{f^2H^{D-2}}\frac{1}{H}, 
\end{align}
where $t_0=\frac{1}{H}\log a_0$.  
Therefore, substituting $\log (a(t)/a_0)\sim 1/(f^2H^{D-2})$, we can estimate the saturation value as 
\begin{align}
\langle (\varphi^i(x)\varphi^i(x))^n \rangle
\sim H^{n(D-2)}\cdot \frac{1}{(f^2H^{D-2})^n}=\frac{1}{f^{2n}}\hspace{1em}\text{at $t-t_0\gg t_c-t_0$}. 
\label{estimate2}\end{align}

We also mention the flat space limit of the IR effect. 
Although the saturation value (\ref{example}) itself does not depend on the Hubble constant, 
this fact does not mean that this calculation result also holds true in flat space. 
As discussed above, 
$\langle (\varphi^i(x)\varphi^i(x))^n \rangle$ approaches the saturation value at $t-t_0 \gg t_c-t_0$. 
This time region disappears in the flat space limit $H\to 0$. 
Therefore, we should take the flat space limit of (\ref{estimate1}) rather than that of (\ref{estimate2}). 
The flat space limit of (\ref{estimate1}) shows the expected result that there is no IR effect in flat space.   

\section{More general model}
\setcounter{equation}{0}

As a more general model, we consider the hybrid model 
which consists of a noncanonical kinetic term and a potential term:  
\begin{align}
S=\int \sqrt{-g}d^Dx\ 
\left[-\frac{1}{2}G_{ij}(\varphi)g^{\mu\nu}\partial_\mu\varphi^i\partial_\nu\varphi^j-V(\varphi)\right]. 
\label{Hybrid}\end{align}
We consider the case that each coupling in the potential term, 
which is made dimensionless by the Hubble constant, is kept small. 
In other words, the scalar fields can be identified as pseudo Nambu-Goldstone bosons. 

The field equation for the hybrid model is written in the covariant form 
\begin{align}
\frac{1}{\sqrt{-g}}\partial_\mu\big\{\sqrt{-g}\ E_i^{\ a}(\varphi)g^{\mu\nu}\partial_\nu\varphi^i\big\}-D_iE_j^{\ a}(\varphi)g^{\mu\nu}\partial_\mu\varphi^i\partial_\nu\varphi^j
-E^{ia}(\varphi)D_iV(\varphi)=0, 
\label{FE2}\end{align}
and thus the Yang-Feldman equation is given by 
\begin{align}
-i\int d^D&x'\ \partial'_\mu\left\{\sqrt{-g(x')}g^{\mu\nu}(x')\partial'_\nu G^R(x,x') E_i^{\ a}(\varphi(x'))\right\}
\varphi^i(x') \notag\\
&=\varphi_0^a(x)
-i\int\sqrt{-g(x')}d^Dx'\ G^R(x,x')\left\{D_iE_j^{\ a}(\varphi(x'))
g^{\mu\nu}(x')\partial'_\mu\varphi^i(x')\partial'_\nu\varphi^j(x')\right. \notag\\
&\hspace{17.5em}\left.+E^{ia}(\varphi(x'))D_iV(\varphi(x'))\right\}.  
\label{YF2}\end{align}

Applying the leading logarithm approximation (\ref{right1}), (\ref{right2}), (\ref{right3}) and (\ref{left}), 
the Yang-Feldman equation reduces to 
\begin{align}
E_i^{\ a}(\varphi(x))\varphi^i&(x)
-\int^tdt'\ \partial'_tE_i^{\ a}(\varphi(t',\textbf{x}))\varphi^i(t',\textbf{x}) \notag\\
&=\bar{\varphi}_0^a(x)
+\frac{H^{D-1}}{(4\pi)^\frac{D}{2}}\frac{\Gamma(D-1)}{\Gamma(\frac{D}{2})}
\int^tdt'\ D_iE^{ia}(\varphi(t',\textbf{x})) \notag\\
&\hspace{3.8em}-\frac{1}{(D-1)H}\int^tdt'\ E^{ia}(\varphi(t',\textbf{x}))D_iV(\varphi(t',\textbf{x})). 
\end{align}
Differentiating both sides with respect to $t$, we obtain the Langevin equation with white noise: 
\begin{align}
E_i^{\ a}(\varphi(x))\dot{\varphi}^i(x)=\dot{\bar{\varphi}}_0^a(x)
&+\frac{H^{D-1}}{(4\pi)^\frac{D}{2}}\frac{\Gamma(D-1)}{\Gamma(\frac{D}{2})}D_iE^{ia}(\varphi(x)) \notag\\
&-\frac{1}{(D-1)H}E^{ia}(\varphi(x))D_iV(\varphi(x)), 
\label{Langevin2}\end{align}
\begin{align}
\langle\dot{\bar{\varphi}}_0^a(t,\textbf{x})\dot{\bar{\varphi}}_0^b(t',\textbf{x})\rangle
=\frac{2H^{D-1}}{(4\pi)^\frac{D}{2}}\frac{\Gamma(D-1)}{\Gamma(\frac{D}{2})}\delta^{ab}\delta(t-t'). 
\label{white2}\end{align}
Here we explain why the Langevin equation for the hybrid model can be derived 
in the same way as that for the nonlinear sigma model. 
Since (\ref{right1}), (\ref{right3}) and (\ref{left}) are the approximations of the free field $\varphi_0$, 
we can apply them regardless of the type of interactions. 
Although (\ref{right2}) is the approximation of the interacting field $\varphi$, 
it holds true regardless of whether the potential term is present or not.  
Since the differential operators in (\ref{right2}) reduce the number of IR logarithms 
from nonderivative interactions, 
(\ref{right2}) does not depend on the potential term up to the leading logarithm accuracy. 
As discussed in (\ref{diagrams}), we can keep up the number of IR logarithms from derivative interactions 
by constructing a loop of twice-differentiated propagators. 

From the Langevin equation in (\ref{Langevin2}) and (\ref{white2}), we can derive the Fokker-Planck equation: 
\begin{align}
\frac{\dot{\rho}(t,\phi)}{\sqrt{G(\phi)}}
&=\frac{H^{D-1}}{(4\pi)^\frac{D}{2}}\frac{\Gamma(D-1)}{\Gamma(\frac{D}{2})}
D^iD_i\left\{\frac{\rho(t,\phi)}{\sqrt{G(\phi)}}\right\}
+\frac{1}{(D-1)H}D^i\left\{D_iV(\phi)\frac{\rho(t,\phi)}{\sqrt{G(\phi)}}\right\} \notag\\
&=\frac{H^{D-1}}{(4\pi)^\frac{D}{2}}\frac{\Gamma(D-1)}{\Gamma(\frac{D}{2})}
D^i\left\{\left[D_i+\frac{2\pi^\frac{D+1}{2}}{\Gamma(\frac{D+1}{2})H^D}D_iV(\phi)\right]
\frac{\rho(t,\phi)}{\sqrt{G(\phi)}}\right\}. 
\label{FP2}\end{align}
In order to solve the Fokker-Planck equation for an equilibrium state, it is useful to introduce 
\begin{align}
\Psi(t,\phi)=\exp\left\{\frac{1}{2}\frac{2\pi^\frac{D+1}{2}}{\Gamma(\frac{D+1}{2})H^D}V(\phi)\right\}
\frac{\rho(t,\phi)}{\sqrt{G(\phi)}}. 
\end{align}
We can rewrite (\ref{FP2}) as the equation for the rescaled probability density:    
\begin{align}
\dot{\Psi}(t,\phi)=-\frac{H^{D-1}}{(4\pi)^\frac{D}{2}}\frac{\Gamma(D-1)}{\Gamma(\frac{D}{2})}
\mathcal{A}^{\dagger i} \mathcal{A}_i\Psi(t,\phi), 
\end{align}
\begin{align}
\mathcal{A}_i&=D_i+\frac{1}{2}\frac{2\pi^\frac{D+1}{2}}{\Gamma(\frac{D+1}{2})H^D}D_iV(\phi), \notag\\
\mathcal{A}^{\dagger i}&=-D^i+\frac{1}{2}\frac{2\pi^\frac{D+1}{2}}{\Gamma(\frac{D+1}{2})H^D}D^iV(\phi). 
\end{align}
Since $\mathcal{A}^{\dagger i} \mathcal{A}_i$ is non-negative--definite, 
the solution for the equilibrium sate, $\Psi(t,\phi)\to \Psi_\infty(\phi)$ at $t\to\infty$, is given by 
\begin{align}
\mathcal{A}_i\Psi_\infty(\phi)=0\hspace{1em}\Rightarrow\hspace{1em}
\rho_\infty(\phi)=Z^{-1}
\sqrt{G(\phi)}\exp\left\{-\frac{2\pi^\frac{D+1}{2}}{\Gamma(\frac{D+1}{2})H^D}V(\phi)\right\}, 
\label{solution2}\end{align}
where the normalization factor is given by 
\begin{align}
Z=\int d^N\phi\ 
\sqrt{G(\phi)}\exp\left\{-\frac{2\pi^\frac{D+1}{2}}{\Gamma(\frac{D+1}{2})H^D}V(\phi)\right\}. 
\label{solution2'}\end{align} 
If $V(\phi)$ behaves like $\lambda|\phi|^\alpha,\ \lambda,\ \alpha>0$ at $|\phi|\to \infty$, 
even though the asymptotic behavior of $\sqrt{G(\phi)}$ is at most $|\phi|^{-\beta}$, $\beta>0$, 
an equilibrium state is eventually reached. 

The equilibrium solution in (\ref{solution2}) and (\ref{solution2'}) indicates a relation 
to the Euclidean field theory on a $D$-dimensional sphere. 
The IR effects discussed in this paper are interpreted as nonequilibrium phenomena 
as they break the dS symmetry and evolve with time. 
However, if an equilibrium state is eventually reached, 
it may be described by the Euclidean field theory on a $D$-dimensional sphere. 
For nonderivative interactions, it has been known that the equilibrium solution of the stochastic equation 
can be reproduced by considering the zero mode dynamics in the Euclidean field theory 
on a $D$-dimensional sphere \cite{Hu1986,Hu1987,Rajaraman2010}. 
As shown below, the discussion holds true even when the kinetic term is noncanonical. 

The scalar fields can be expanded by the spherical harmonics $Y_{\{l_1,\cdots,l_D\}}$ 
on a $D$-dimensional sphere as 
\begin{align}
\varphi^i(x_E)=\sum_{\{l_1,\cdots,l_D\}}\varphi^i_{\{l_1,\cdots,l_D\}}Y_{\{l_1,\cdots,l_D\}}(x_E), 
\end{align}
where $x_E$ are the spherical coordinates 
and $l_1\ge |l_2| \ge \cdots \ge |l_D|\ge 0$ are the angular momentums. 
In order to evaluate the IR effects, we extract the zero mode as 
\begin{align}
\varphi^i(x_E)\simeq \varphi^i_{\{0,\cdots,0\}}Y_{\{0,\cdots,0\}}. 
\end{align}
It should be recalled that the zero mode has no coordinate dependence. 

In the zero mode approximation, only the potential term contributes to the Euclidean action as 
\begin{align}
S_E
&=\int\sqrt{g_E}d^Dx_E\ \left[\frac{1}{2}G_{ij}(\varphi)g_E^{\mu\nu}\partial_\mu\varphi^i\partial_\nu\varphi^j
+V(\varphi)\right] \notag\\
&\simeq V(\varphi_{\{0,\cdots,0\}}Y_{\{0,\cdots,0\}})\int\sqrt{g_E}d^Dx_E \notag\\
&=V(\varphi_{\{0,\cdots,0\}}Y_{\{0,\cdots,0\}})\cdot \frac{2\pi^\frac{D+1}{2}}{\Gamma(\frac{D+1}{2})H^D}, 
\label{SE}\end{align}
where the coefficient of the potential term is nothing but the volume 
of a $D$-dimensional sphere of radius $1/H$. 
In the path integral formalism with (\ref{SE}), the vev of any operator made of $\varphi^i$ is evaluated as 
\begin{align}
&\langle F^{i_1\cdots i_n}(\varphi(x))\rangle \label{vev'}\\
=\ &\frac{\int \sqrt{\mathcal{G}}\mathcal{D}\varphi\ F^{i_1\cdots i_n}(\varphi(x))e^{-S_E}}
{\int \sqrt{\mathcal{G}}\mathcal{D}\varphi\ e^{-S_E}} \notag\\
\simeq\ &\frac{\int\sqrt{G}d^N(\varphi_{\{0,\cdots,0\}}Y_{\{0,\cdots,0\}})\ 
F^{i_1\cdots i_n}(\varphi_{\{0,\cdots,0\}}Y_{\{0,\cdots,0\}})
\exp\left\{-\frac{2\pi^\frac{D+1}{2}}{\Gamma(\frac{D+1}{2})H^D}
V(\varphi_{\{0,\cdots,0\}}Y_{\{0,\cdots,0\}})\right\}}
{\int\sqrt{G}d^N(\varphi_{\{0,\cdots,0\}}Y_{\{0,\cdots,0\}})\ 
\exp\left\{-\frac{2\pi^\frac{D+1}{2}}{\Gamma(\frac{D+1}{2})H^D}
V(\varphi_{\{0,\cdots,0\}}Y_{\{0,\cdots,0\}})\right\}}. \notag
\end{align}
The covariant functional integral $\sqrt{\mathcal{G}}\mathcal{D}\varphi$ reduces to 
$\sqrt{G}d^N(\varphi_{\{0,\cdots,0\}}Y_{\{0,\cdots,0\}})$ in the zero mode approximation. 
Identifying $\varphi_{\{0,\cdots,0\}}Y_{\{0,\cdots,0\}}$ as $\phi$, 
it can be confirmed that (\ref{vev'}) is equivalent to the stochastic evaluation 
with the equilibrium solution in (\ref{solution2}) and (\ref{solution2'}). 

\section{Conclusion}
\setcounter{equation}{0}

We extended the resummation formula of the leading IR logarithms 
so that it is applicable to a general scalar field theory whose kinetic term is noncanonical. 
In the presence of derivative interactions, 
we need to take into account not only the contribution from the superhorizon scale 
but also that from the subhorizon scale. 
The discussion around (\ref{vertex})--(\ref{right2}) shows that up to the leading logarithm accuracy, 
we can integrate out differentiated fields as if nondifferentiated fields were constant. 
Specifically, the subhorizon dynamics can be treated as in the free field theory 
with the coefficient of the kinetic term $G_{ij}(\varphi)$. 
After integrating out the contribution from the subhorizon scale, 
we treat nondifferentiated fields as dynamical variables. 

In the general scalar field theory, 
the leading logarithm approximation of the Yang-Feldman equation 
leads to a Langevin equation with white noise.   
As seen in (\ref{FP2}), the resulting stochastic equation can be understood 
as a generalization of the standard one to curved target spaces. 
The generalized stochastic equation shows that if the target space is sufficiently compact as in (\ref{compact}), 
even though the potential term is absent, an equilibrium state is eventually reached.  
The equilibrium state can be reproduced by considering the zero mode dynamics 
in the Euclidean field theory on a sphere. 
It should be emphasized that if the target space is not so compact,  
i.e. the asymptotic behavior of $\sqrt{G(\phi)}$ is at most $|\phi|^{-\beta}$, $\beta>0$,    
and the potential term is absent, no equilibrium state is reached. 

We derived the IR resummation formula for the general scalar field theory 
as a first step to obtain that for gravity. 
The Einstein gravity consists of derivative interactions 
in a similar way to the general scalar field theory\footnote{
More specifically, neither the Einstein gravity nor the models adopted in this paper 
include higher derivative interactions. }
and includes gauge degrees of freedom and tensor fields in contrast to it. 
This study shows how to resum the leading IR logarithms from derivative interactions of the scalar field. 
The resummation formula in the presence of gauge degrees of freedom and tensor fields is 
a subject for future research.  

The contribution from gravity should be considered also in evaluating the IR effects in inflation theories. 
If only the inflaton induced the IR effects, we could evaluate them nonperturbatively 
by using the current resummation formula. 
However, there is no reason to neglect the IR effects from gravity.   
In particular, if the pseudo shift symmetry is respected, 
the IR effects from the gravitational interaction are more dominant 
compared with those from the self-interaction of the inflaton \cite{Kitamoto2017}. 

\section*{Acknowledgment}

This work was supported in part by the National Center of Theoretical Sciences (NCTS) 
and Grant-in-Aid for Scientific Research (B) No. 26287044. 
We thank G. Cho, C. S. Chu, K. Furuuchi, B. L. Hu, C. H. Kim, Y. Kitazawa, S. P. Miao, S. Nojiri, T. Tanaka, 
and R. P. Woodard for discussions. 
In particular, we thank G. Cho and C. H. Kim for their participation at an early stage of this work, 
and K. Furuuchi for reading the manuscript and making comments. 

\appendix

\section{Langevin and Fokker-Planck equations}
\setcounter{equation}{0}

Here we review the relation between the Langevin equation and the Fokker-Planck equation. 
For a derivation, see, e.g., \cite{Risken}.  

If the Langevin equation is expressed as 
\begin{align}
\dot{\xi}^i(t)=A^i_{\ a}(t,\xi(t))\eta^a(t)+B^i(t,\xi(t)), 
\label{Langevin0}\end{align}
\begin{align}
\langle \eta^a(t)\rangle=0,\hspace{1em}
\langle \eta^a(t)\eta^b(t')\rangle=\delta^{ab}\delta(t-t'). 
\label{white0}\end{align}
The corresponding Fokker-Planck equation is given by 
\begin{align}
\dot{\rho}(t,x)=
\ &\frac{1}{2}\frac{\partial^2}{\partial x^i\partial x^j}\left\{A^i_{\ a}(t,x)A^{ja}(t,x)\rho(t,x)\right\} \notag\\
&-\frac{\partial}{\partial x^i}\left\{\left[\frac{1}{2}\frac{\partial}{\partial x^j}A^i_{\ a}(t,x) A^{ja}(t,x)
+B^i(t,x)\right]\rho(t,x)\right\}. 
\label{FP0}\end{align}
Using its solution, the vev of any operator made of $\xi^i$ can be evaluated as  
\begin{align}
\langle F^{i_1\cdots i_n}(\xi(t))\rangle=\int d^Nx\ F^{i_1\cdots i_n}(x)\rho(t,x), 
\label{F0}\end{align}
where $F^{i_1\cdots i_n}$ denotes an arbitrary function. 
In deriving (\ref{FP1}) from (\ref{Langevin1}) and (\ref{white1}), we made the following replacement: 
\begin{align}
&\xi^i(t)\to \varphi^i(t,\textbf{x}), \notag\\
&A^i_{\ a}\to 
\left\{\frac{2H^{D-1}}{(4\pi)^\frac{D}{2}}\frac{\Gamma(D-1)}{\Gamma(\frac{D}{2})}\right\}^\frac{1}{2}
E^i_{\ a}, \notag\\
&B^i\to \frac{H^{D-1}}{(4\pi)^\frac{D}{2}}\frac{\Gamma(D-1)}{\Gamma(\frac{D}{2})}E^i_{\ a}D_jE^{ja}. 
\end{align}

In the main text, the probability density for one-point functions is discussed. 
Here we study the probability density for spatially separated two-point functions.  
In order to study it, we need to know the correlation function of $\dot{\bar{\varphi}}_0^a$ 
at spatially separated points: 
\begin{align}
\langle\dot{\bar{\varphi}}_0^a(t,\textbf{x})\dot{\bar{\varphi}}_0^b(t',\textbf{x}')\rangle
&=\frac{2H^{D-1}}{(4\pi)^\frac{D}{2}}\frac{\Gamma(D-1)}{\Gamma(\frac{D}{2})}\delta^{ab}\delta(t-t')
\theta(1-Ha(t)r), 
\label{white1'}\end{align}
where $r=|\textbf{x}-\textbf{x}'|$. 
In deriving (\ref{white1'}), 
we approximated the Bessel function of the first kind by the step function as 
\begin{align}
\frac{2^\frac{D-3}{2}\Gamma(\frac{D-1}{2})}{(Ha(t)r)^\frac{D-3}{2}}J_\frac{D-3}{2}(Ha(t)r)
\simeq \theta(1-Ha(t)r).  
\end{align}

From (\ref{Langevin1}) and (\ref{white1'}), 
the Fokker-Planck equation for spatially separated two-point functions is given by 
\begin{align}
\frac{\dot{\rho}(t,\phi,\phi')}{\sqrt{G(\phi)G(\phi')}}=
\ &\frac{H^{D-1}}{(4\pi)^\frac{D}{2}}\frac{\Gamma(D-1)}{\Gamma(\frac{D}{2})}
D^iD_i\left\{\frac{\rho(t,\phi,\phi')}{\sqrt{G(\phi)G(\phi')}}\right\} \\
&+\frac{H^{D-1}}{(4\pi)^\frac{D}{2}}\frac{\Gamma(D-1)}{\Gamma(\frac{D}{2})}
D'^iD'_i\left\{\frac{\rho(t,\phi,\phi')}{\sqrt{G(\phi)G(\phi')}}\right\} \notag\\
&+\theta(1-Ha(t)r)\frac{2H^{D-1}}{(4\pi)^\frac{D}{2}}\frac{\Gamma(D-1)}{\Gamma(\frac{D}{2})}
D_i\left\{E^i_{\ a}(\phi)D'_j\left[E^{ja}(\phi')\frac{\rho(t,\phi,\phi')}{\sqrt{G(\phi)G(\phi')}}\right]\right\}, \notag
\end{align}
where $\phi$ and $\phi'$ denote $\phi(\textbf{x})$ and $\phi(\textbf{x}')$. 
Using its solution, any spatially separated two-point function can be evaluated as 
\begin{align}
\langle P^{i_1\cdots i_m}(\varphi(t,\textbf{x}))Q^{j_1\cdots j_n}(\varphi(t,\textbf{x}'))\rangle
=\int d^N\phi  d^N\phi'\ P^{i_1\cdots i_m}(\phi)Q^{j_1\cdots j_n}(\phi')\rho(t,\phi,\phi'), 
\end{align}
where $P^{i_1\cdots i_m}$ and $Q^{j_1\cdots j_n}$ denote arbitrary functions. 


\end{document}